\documentclass[prd,preprint,nofootinbib]{revtex4}

\usepackage{graphicx}
\usepackage{multirow}
\usepackage{subfigure}
\usepackage{amssymb,amsmath}
\usepackage{color}
\usepackage{enumitem}
\usepackage{braket}
\usepackage{array}
\usepackage{setspace}
\usepackage{ulem}
\usepackage{hyperref}

\newcommand{\Nuniv}{N_{\rm univ.}}

\newcolumntype{x}[1]{>{\centering\arraybackslash\hspace{0pt}}p{#1}}

\begin{document}
	\title{Revisiting leptonic nonunitarity}

	\author{Daniel Aloni}
	\email{alonidan@bu.edu}
	\affiliation{Physics Department, Boston University, Boston, Massachusetts 02215, USA}
	\affiliation{Department of Physics, Harvard University, Cambridge, Massachusetts 02138, USA}
	\author{Avital Dery}
	\email{avital.dery@cornell.edu}
	\affiliation{Department of Physics, LEPP, Cornell University, Ithaca, New York 14853, USA}

	\begin{abstract}
		In the presence of extra neutrino states at high scales, the low-energy effective $3\times 3$ leptonic mixing matrix (LMM) is in general nonunitary. We revisit the question of what is our current knowledge of individual LMM matrix elements without assuming unitarity. 
		We first demonstrate that a minimal set of experimental constraints suffices in bounding LMM nonunitarity parameters to the level of ${\cal O}(10^{-3})$, without the use of neutrino oscillation data.
		We then revisit oscillation results as a complementary cross-check, using different physics and different experimental techniques to probe a similar parameter space. 
		We correct some common misconceptions in the neutrino nonunitarity literature resulting from an incautious treatment of input parameters. We find that neutrino oscillation experiments can constrain LMM nonclosure, but, contrary to claims in the literature, are completely insensitive to the overall normalization of the LMM. Thus we conclude that oscillation experiments, including the future DUNE experiment, have no power in excluding nonunitarity altogether.
	\end{abstract}
	\maketitle

\section{Introduction}\label{sec:intro}
%
The discovery of neutrino oscillations is a milestone in the history of particle physics. 
This phenomenon was the first and so far only direct observation of physics beyond the standard model (BSM). 
Neutrino oscillations are a consequence of ($i$) at least two neutrinos being massive, and ($ii$) mass eigenstates being misaligned with weak interaction eigenstates.
While these properties are absent in the SM, in the minimally required extensions only two mass differences and a $3\times 3$ unitary mixing matrix need to be introduced.
The unitary 'Pontecorvo-Maki-Nakagawa-Sakata' (PMNS) mixing matrix has three angles and one Dirac phase.\footnote{If the neutrinos are Majorana, there exist two additional Majorana phases, but these do not affect oscillation probabilities.}
Since the first observation of missing neutrinos at the Homestake experiment~\cite{Cleveland:1998nv} a grand experimental program succeeded in measuring the two mass squared differences and the three angles to the level of a few percent~\cite{Esteban:2020cvm}.
Nevertheless, even in the minimal scenario many questions remain open. Those are the overall scale and the neutrino mass ordering, the Majorana or Dirac nature, and whether or not \textit{CP} symmetry is broken in the leptonic sector.

In a huge class of models leading to neutrino masses and mixing, new sterile states need to be introduced. These are Weyl fermions which are singlets under the SM gauge group. In minimal scenarios these states are either very heavy and give mass to SM neutrinos via the seesaw mechanism (Majorana neutrino), or very light and generate the neutrino masses via the Higgs mechanism (Dirac neutrino). Nevertheless, it is plausible to introduce additional neutrino states which might lead to richer phenomena of neutrino oscillations (see e.g., Ref.~\cite{ParticleDataGroup:2020ssz} and references therein). 

In such nonminimal scenarios, if the mass scale of the additional neutrino states lies above the kinematic reach of a considered experiment, the consequences of their existence at low energies are embodied in the nonunitarity of the effective low-energy $3\times 3$ leptonic mixing matrix (LMM). In the most general case, the LMM consists of nine magnitudes and four phases, in contrast to the unitary PMNS where three magnitude and one phase suffice. 

In the past two decades a huge effort has been carried out in order to study deviations from unitarity of the LMM. Since Ref.~\cite{Czakon:2001em} first explored the topic, and the seminal paper of Antusch \textit{et al.}~\cite{Antusch:2006vwa} put the theoretical grounds for the formalism of nonunitary neutrino oscillations, many authors confront different oscillation data with the theoretical predictions~\cite{Goswami:2008mi,Antusch:2009pm,Qian:2013ora,Parke:2015goa,Blennow:2016jkn,Escrihuela:2016ube,Fong:2016yyh,DeGouvea:2019kea,Forero:2021azc,Denton:2021mso}. Throughout the years, different parametrizations of the LMM matrix were presented, with the well-known $\eta$ parametrization of Ref.~\cite{Fernandez-Martinez:2007iaa}, and the $\alpha$ parametrization of Ref.~\cite{Escrihuela:2015wra}. Recently Ref.~\cite{Ellis:2020hus} presented yet another convention by packing the information in terms of normalization and nonclosure factors, which have a clear physical interpretation and reveal different aspects of nonunitarity physics. On top of the information extracted from oscillations, Refs.~\cite{Abada:2013aba,Antusch:2014woa,Fernandez-Martinez:2016lgt} extensively studied bounds from nonoscillation-related constraints, and Ref.~\cite{Coutinho:2019aiy} preformed a global fit to all data.

The experimental constraints are traditionally split in the literature between bounds from neutrino experiments compared to an extensive set of complementary measurements.
The former involves independent measurements of all nine matrix elements and four phases, which in general is highly challenging. It is traditionally unjustifiedly motivated by the claim that, unlike other constraints, constraints from this dataset are model independent.

Current constraints on LMM nonunitarity arise from a variety of measurements, including, for example: electroweak data, universality tests, radiative lepton decays and oscillation data. 
In this work we do not aim to perform a global fit to these constraints, but rather study the implications and limitations of individual constraints by splitting the collection of relevant measurements into two minimal complementary sets:
\begin{enumerate}
	\item {\it Minimal and robust} --- We demonstrate that a minimal set of (i) lepton flavor universality (LFU) ratios and (ii) weak angle measurements, is able to place stringent bounds on nonunitarity parameters.
	
	\item {\it Oscillation  $+$ LFU} --- We treat a dataset of oscillation data together with the LFU bounds as a complementary cross-check of LMM nonunitarity constraints. We argue that using oscillation data without using the LFU constraints is inconsistent.
\end{enumerate}
In studying our second dataset we draw special attention to contamination in input parameters and fix some common errors in the literature. We provide the correct expressions which are needed to consistently analyze neutrino oscillation data in the framework of nonunitarity, and show that a correct treatment of input parameters completely removes the ability of oscillation experiments to exclude LMM nonunitarity. This is in sharp contradiction to current literature. We emphasize that neutrino oscillation experiments are sensitive only to the nonclosure of the mixing matrix, and hence that should be the only target of such experiments. %

The structure of the paper is as follows: In Sec.~\ref{sec:formalism} we review the formalism needed for nonunitarity of the LMM. In Sec.~\ref{sec:minimal} we update the bounds arising from LFU and weak mixing angle measurements and derive the resulting constraints on nonunitarity. In Sec.~\ref{sec:complementary} we study the (in)sensitivity of neutrino oscillation data, discussing the subtleties and misconceptions in the literature. %
We conclude in Sec.~\ref{sec:conclusion}. Complementary material is given in Appendixes~\ref{app:oscillation probabilities} and \,\ref{app:LFU}\,.

\section{Formalism}
\label{sec:formalism}
In this section we review the framework and formalism used throughout this paper. We mainly follow the notations of Ref.~\cite{Ellis:2020hus}, while we clarify explicitly in places where we differ. The formalism is based on Ref.~\cite{Giunti:2004zf} which was developed to the scenario of nonunitary LMM in Ref.~\cite{Antusch:2006vwa}.
We consider models in which extra sterile neutrino states exist at scales above the relevant scale for neutrino production, and whose consequences at low energies are embodied in the nonunitarity of the $3\times 3$ leptonic mixing matrix (LMM).

In its most general form the LMM contains nine magnitudes and four phases, and can always be rotated to the following form,
\begin{align}\label{eq: pmns matrix}
	U & = \begin{pmatrix}
		\left|U_{e 1}\right| &  \left|U_{e 2}\right|e^{i\phi_{e 2}} &  \left|U_{e 3}\right| e^{i\phi_{e 3}}\\
		\left|U_{\mu 1}\right| &  \left|U_{\mu 2}\right| &  \left|U_{\mu 3}\right| \\
		\left|U_{\tau 1}\right| &  \left|U_{\tau 2}\right|e^{i\phi_{\tau 2}} &  \left|U_{\tau 3}\right| e^{i\phi_{\tau 3}}
	\end{pmatrix} ~.
\end{align}

The effective neutrino flavor state produced in an experiment in association with a charged lepton, $\ell_\alpha$, from a hadron or lepton decay at $t=0$ after propagating in vacuum for a time $t$, is given by
\begin{align}\label{eq: ket neutrino state}
	\ket{\nu_\alpha (t)^{\rm eff.}} & = \frac{1}{\sqrt{N_{\alpha}}}\sum_{k=1}^{3} U^*_{\alpha k} e^{i E_k t} \ket{\nu_k}~,
\end{align}
where $\alpha = e,\,\mu,\,\tau$ stands for the neutrino flavor and $k$ stands for the neutrino mass eigenstate. The sum is truncated at three, as under our assumptions mass eigenstates with $k>3$ are kinematically inaccessible at production. The normalization factor $N_\alpha$  is chosen such that $\braket{\nu_\alpha (t)| \nu_\alpha(t)} = 1$.
In general we define the normalization factors,
\begin{align}
	N_\alpha & \equiv \left|U_{\alpha 1}\right|^2 + \left|U_{\alpha 2}\right|^2 + \left|U_{\alpha 3}\right|^2 \qquad,\qquad (\alpha = e,\,\mu,\,\tau) ~, \\
	N_k & \equiv \left|U_{e k}\right|^2 + \left|U_{\mu  k}\right|^2 + \left|U_{\tau k}\right|^2 \qquad,\qquad (k = 1,\,2,\,3) ~, 
\end{align}
and the nonclosure factors,
\begin{align}\label{eq:nonclosure}
	t_{\alpha\beta} & \equiv U^*_{\alpha 1} U_{\beta 1}  +   U^*_{\alpha 2} U_{\beta 2}  +U^*_{\alpha 3} U_{\beta 3} ~, \\
	t_{k l} &  \equiv U^*_{e k} U_{e l}  +   U^*_{\mu k} U_{\mu l}  +U^*_{\tau k} U_{\tau l} ~, 
\end{align}
which are measures of the nonunitarity of the LMM. 
Throughout this work we  assume that the LMM is a submatrix of a larger unitary matrix. That is,
\begin{eqnarray}
	N_i,\, N_\alpha \, \leq \, 1 \qquad i=1,2,3,\quad \alpha=e,\mu,\tau \, .
\end{eqnarray}

The normalization factors and nonclosure factors are not independent; in particular, using the Cauchy-Schwarz inequality on the vectors of the full extended mixing matrix, one can derive  the following relations,
\begin{eqnarray}
	|t_{\alpha\beta}|^2 \, &\leq & \, (1-N_\alpha)(1-N_\beta) \, , \\ \nonumber
	|t_{ij}|^2 \, &\leq & \, (1-N_i)(1-N_j) \, .
\end{eqnarray}
This implies that any nonunitarity exhibited in nonclosure factors must be accompanied by nonunitarity in normalization factors of at least the same magnitude.

\section{A minimal and robust set of constraints}\label{sec:minimal}
%
In this section we demonstrate, that a minimal set of experimental constraints, consisting of: 
\begin{enumerate}
	\item Two lepton flavor universality (LFU) ratios.
	
	\item One ratio of weak angle measurements. 
\end{enumerate} 
is able to exclude deviations from unitarity in the LMM larger than ${\cal O}(10^{-3})$.
These constraints, together with many others, have been considered previously in the literature in the context of LMM unitarity (see, e.g., Ref.~\cite{Antusch:2014woa}). Here we emphasize that a very minimal set such as the one considered here suffices in placing stringent constraints, remarkably without the use of neutrino oscillation data. We update these constraints with the latest relevant experimental data. 

The robustness of these constraints is due to the fact that they are sensitive to nonunitarity at leading order in perturbation theory, and were measured using several independent systems and experimental techniques.
We note that, if the only new physics affecting these measurements is the effective nonunitarity, then the constraints detailed below are unavoidable. 
In order to circumvent them, one would need to alter meson and lepton decays, introducing additional couplings to light and invisible states (to avoid LFU constraints), or introduce effective operators that affect weak-angle determinations (to avoid EW constraints).

\subsection{Lepton flavor universality bounds}
\label{sec:LFU}
One of the most robust constraints on row normalization factors can be derived from non-universality constraints in meson and lepton decays (see Refs~\cite{Antusch:2006vwa,Abada:2013aba,Antusch:2014woa}).
Here, a ratio of branching ratios of identical processes with a single different flavor of charge lepton is proportional to the ratio of normalization factors. 
For instance one finds the following from LFU measurements of pion decay (for a detailed derivation see Appendix~\ref{app:LFU}):
\begin{align}
	\frac{{\rm BR}(\pi \to \mu \nu)}{{\rm BR}(\pi \to e \nu)} &  = \left(\frac{{\rm BR}(\pi \to \mu \nu)}{{\rm BR}(\pi \to e \nu)} \right)^{\rm SM} \frac{N_\mu}{N_e}~.
\end{align}
LFU constraints arise from tree-level processes and are applicable in any model in which the extra states are kinematically inaccessible at meson and lepton mass scales.
In order to avoid the LFU bounds one has to introduce either light fermions which are singlets under the SM gauge group, also known as light sterile neutrinos, and/or introduce nonstandard neutrino interactions in a way that mimics the SM prediction. Therefore we argue these limits cannot be avoided without further NP in the neutrino sector.

In our analysis, we use a set of LFU constraints provided by the HFLAV Collaboration~\cite{HFLAV:2019otj}, as summarized in Table~\ref{tab:LFU_HFLAV}. More details as to LFU bounds from additional processes can be found in Appendix~\ref{app:LFU}.

The conjunction of any two ratios from Table~\ref{tab:LFU_HFLAV} leads to the conclusion that all row normalization factors, 
$N_\alpha$, can be thought of as moving together in the allowed flat direction, up to deviations of ${\cal O}(10^{-3})$ at most. 
The only considerable freedom that remains is of the universal normalization, which we denote as $\Nuniv$,
\begin{equation}
	N_e \simeq N_\mu \simeq N_\tau \simeq \Nuniv\, .
\end{equation}

\begin{table}[t]
	\begin{center}
		\bgroup
		\def\arraystretch{1.5}
		\begin{tabular}{|c||c |c| c |}
			\hline
			Ratio &  Bound & Processes\\ 
			\hline \hline
			$\frac{N_\mu}{N_e}$ & $1.004 \pm 0.003$ & lepton decays \\
			\hline
			$\frac{N_\tau}{N_e}$ & $1.006 \pm 0.003$ & lepton decays \\
			\hline
			$\frac{N_\tau}{N_\mu}$ & $0.999 \pm 0.003$ & lepton, $	\pi,\, K$ decays \\
			\hline
		\end{tabular}
		\egroup
		\caption{Summary of  LFU constraints on ratios of normalization factors, taken from Ref.~\cite{HFLAV:2019otj}.}
		\label{tab:LFU_HFLAV}
	\end{center}
\end{table}

\subsection{Weak-mixing angle bounds}
\label{sec:EW}
As was shown in Ref.~\cite{Antusch:2014woa}, the combination of various weak mixing angle measurements places a strong bound on the universal normalization, $\Nuniv$. In the following we update the experimental inputs to this derivation. The Fermi constant as measured from muon decay is sensitive to non unitarity of the LMM matrix at tree level
\begin{equation}\label{eq: Gmu}
	G_\mu \, = \, G_F \sqrt{N_e N_\mu}\,.
\end{equation}
This measurement can be compared to the value of $\sin^2\theta_{W}$ measured by using other systems~\cite{Antusch:2014woa,Crivellin:2021njn}, and can be interpreted as setting stringent bounds on the product $N_e N_\mu$. 
We consider leptonic $Z$-boson decays, and update the effective weak mixing angle estimations using the latest LHC data.

The effective weak mixing angle, $\sin^2{\theta_{\rm eff}^f}$~\cite{Degrassi:1990ec,Gambino:1993dd}, is measured in $Z$ boson decays to fermion pairs, via forward-backward asymmetries or polarization asymmetries. Comparing the SM theoretical prediction~(see Sec.~10.5 of~\cite{ParticleDataGroup:2020ssz}) with the expectation of a nonunitary LMM, one finds
\begin{equation}
	\frac{\left(\sin^2{\theta_{\rm eff}^f}\cos^2{\theta_{\rm eff}^f}\right)_{\rm SM}}{\left(\sin^2{\theta_{\rm eff}^f}\cos^2{\theta_{\rm eff}^f}\right)_{\rm NU}} \, =\sqrt{N_e N_\mu}\, . 
\end{equation}
where the subscript $\rm NU$ stands for the nonunitary case.
Here we update the results of Ref.~\cite{Antusch:2014woa} by combining results from the Tevatron CDF and D\O{} experiments~\cite{CDF:2016cei,D0:2017ekd}, and the LHC ATLAS, CMS and LHCb experiments~\cite{ATLAS:2015ihy,ATLAS:2018gqq,CMS:2018ktx,LHCb:2015jyu}.  Following~\cite{ParticleDataGroup:2020ssz} for combining leptonic ($\ell = e,\mu$) observables, we have
\begin{equation}
	\sin^2{\theta_{\rm eff}^\ell}\cos^2{\theta_{\rm eff}^\ell} \,=\,  0.17784\pm 0.00012\, .
\end{equation}
This is to be compared with the SM prediction~\cite{ParticleDataGroup:2020ssz},
\begin{equation}
	\big(\sin^2{\theta_{\rm eff}^\ell}\cos^2{\theta_{\rm eff}^\ell}\big)_{\rm SM} \,= \, 0.17792 \pm 0.00002\, .
\end{equation}
The resulting constraint reads 
\begin{equation}
	\sqrt{N_e N_\mu} \, = \, 1.0004 \pm 0.0007\, . 
\end{equation}
Together with the nonuniversality constraints of Table~\ref{tab:LFU_HFLAV}, by using simple propagation of error, this implies 
\begin{eqnarray}
	N_e \, &=& \, 0.998 \pm 0.002\, , \qquad
	N_\mu \, = \, 1.002 \pm 0.002\, , \qquad N_\tau \, = \, 1.003 \pm 0.002\, ,
\end{eqnarray}
or in terms of lower bounds,
\begin{eqnarray}
	N_e \, >  \, 0.995 \, , \qquad
	N_\mu \, > \, 0.999 \, , \qquad N_\tau \, > \, 0.998 \qquad \text{at}\,\, 2\sigma \, .
\end{eqnarray}

Thus, without using any input from neutrino experiments, the combination of 
\begin{enumerate}
	\item LFU constraints, and 
	\item Weak mixing angle bounds
\end{enumerate}
suffices in order to exclude deviations of row normalization factors from unity larger than ${\cal O}(10^{-3})$. 

Using the Cauchy-Schwartz inequality, the row nonclosure parameters are similarly constrained,
\begin{equation}
	|t_{\alpha\beta}|^2 \, \leq \, (1-N_\alpha)(1-N_\beta) \, < \, {\cal O}(10^{-6})\, , \qquad \alpha,\,\beta = e,\, \mu, \, \tau\, .
\end{equation}
Finally, as the sum over column normalization factors equals to the sum over row normalization factors, 
\begin{equation}
	\sum_\alpha N_\alpha \, = \, \sum_i N_i \, ,
\end{equation}
we conclude that the sum of column normalization factors is also equal to one at the per mil level. As we take all normalization factors to be at most one, we cannot achieve large cancellations between different column normalization factors. This results in per mil level bounds on each of the column normalization factors as well.

\section{Complementary cross-checks from oscillation data}
\label{sec:complementary}
%
Oscillation experiments are directly sensitive to the LMM parameters.
In this section we work with a dataset excluding the weak mixing angle constraints of the previous section, considering only measurements that directly involve neutrinos in the initial or final state.
We treat oscillation $+$ LFU data in what follows as a complementary cross-check of LMM constraints, and find that, contrary to what is claimed in much of the literature, oscillation experiments are only sensitive to the normalized LMM elements, $\{|U_{\alpha i}|^2/\Nuniv\}$, and have no sensitivity to the overall normalization of the LMM rows and columns.
As discussed in Sec.~\ref{sec:LFU}, LFU constraints cannot be avoided without modifying neutrino physics beyond nonunitarity.
Moreover, some of the LFU bounds arise from the same meson decays that play a role in the production mechanisms of many oscillation experiments.
Hence ignoring LFU constraints but assuming that the only NP in the neutrino sector is embodied in nonunitarity is inconsistent. Therefore, our starting point for the analysis below is the conclusion of the LFU analysis outlined in Sec.~\ref{sec:LFU}.

In order to interpret oscillation data, we derive the oscillation probability expressions in the presence of nonunitarity, and the relation between the number of observed events and these probabilities.

The oscillation probabilities are given by
\begin{align}\label{eq:Prob}
	P(\nu_\alpha \to \nu_\beta,\,t) & = \left|\braket{\nu_\beta^{\rm eff.} |\nu_\alpha(t)^{\rm eff.}}\right|^2~,
\end{align}
where explicit expressions are given in Appendix~\ref{app:oscillation probabilities}. It is important to highlight the effect of neutrino appearance at zero distance, which arises due to the nonorthogonality of the effective flavor states of Eq.~\eqref{eq: ket neutrino state}. The transition probability at zero distance does not vanish but is rather given by
\begin{align}\label{eq: zero distance probability}
	P(\nu_\alpha \to \nu_{\beta},\,0) & = \begin{cases}
		1 & \beta = \alpha \\
		\frac{|t_{\alpha\beta}|^2}{N_\alpha N_\beta} & \beta \neq \alpha
	\end{cases}~.
\end{align}

\subsection{Flux and cross section predictions}
\label{sec:normalization}
In order to relate the probability expressions to a measured number of events in a neutrino experiment, the neutrino flux and its interaction cross section need to be taken into account,
\begin{equation}
	n_\beta \propto \Phi_\alpha \cdot P(\nu_\alpha \to \nu_\beta) \cdot \sigma_\beta\, ,
\end{equation}
where $\Phi_\alpha$ ($\sigma_\alpha$) is the neutrino flux (cross section) associated with a lepton $\ell_\alpha$.
In experiments that use both a far detector (FD) and a near detector (ND) and consider the ratio of measured events (assuming a source of a single neutrino flavor), the flux and cross section cancel in the ratio which is then a clean measurement of the evolution probability.
Conversely, all other experiments require estimations of both the flux and the cross section in order to be interpreted.

Since initially postulated in Ref.~\cite{Antusch:2006vwa}, the leptonic nonunitarity literature traditionally treats the estimation of fluxes in the presence of nonunitarity as differing from their SM expectations by a normalization factor, arising from the effective flavor states of Eq.~\eqref{eq: ket neutrino state}. We argue, however, that the SM prediction is misleading, as input parameters are derived from measurements. Therefore, in reality, no such correction factors need to be applied, since fluxes are estimated using measured branching ratios. 
In the following we provide two characteristic examples of flux estimations in order to convey this point.
\begin{enumerate}
	
	\item Consider a pion source of $\mu$-neutrinos. Given a known pion flux, the $\mu$-neutrino flux is given by
	\begin{equation}
		\Phi_\mu = \Phi_\pi \cdot {\rm BR}(\pi\to\mu\nu)\, .
	\end{equation}
	If the branching ratio is taken from measurement, then there is clearly no need to correct the flux estimation, which would inherently include the effects of nonunitarity.
	In the special case of the pion, the branching ratio can also be estimated without any hadronic input, using only the fully leptonic two-body modes,
	\begin{eqnarray}
		{\rm BR}(\pi\to\mu\nu) &\approx & \frac{  |\braket{\mu\,\nu_\mu^{\rm eff.}|{\cal H}|0}|^2\cdot PS_\mu}{ |\braket{\mu\, \nu_\mu^{\rm eff.}|{\cal H}|0}|^2\cdot PS_\mu+ |\braket{e\,\nu_e^{\rm eff.}|{\cal H}|0}|^2 \cdot PS_e} \, ,
	\end{eqnarray}
	where $PS_{e,\mu}$ stands for the relevant phase-space factor.
	Plugging in the effective flavor states of Eq.~\eqref{eq: ket neutrino state} and the Hamiltonian in the mass basis, we have for each matrix element
	\begin{equation}\label{eq:leptonic matrix element}
		\frac{|\braket{\ell_\alpha \nu_\alpha^{\rm eff.}|{\cal H}|0}|^2}{ |\braket{\ell_\alpha \nu_\alpha|{\cal H}|0}^{\rm SM}|^2} = \left|\frac{1}{\sqrt{N_\alpha}} \sum_{i=1}^3 |U_{\alpha i}|^2\right|^2 =  N_\alpha \, .
	\end{equation}
	Therefore, this results in 
	\begin{eqnarray}
		{\rm BR}(\pi\to\mu\nu) &\approx & \frac{N_\mu \cdot PS_\mu}{N_\mu \cdot PS_\mu + N_e \cdot PS_e} = 1 + {\cal O}\left(\frac{N_e}{N_\mu}\cdot 10^{-4}\right)\, .
	\end{eqnarray}
	In other words, since ${\rm BR}(\pi\to\mu\nu) \approx 1$ is known, nonunitarity cannot result in a non-negligible deviation in the $\mu$-neutrino flux from pions. 
	Similarly, as nuclear reactors operate below the muon threshold and produce only $\nu_e$, no correction should be applied to the flux of reactor neutrinos, which is estimated from the measured produced energy.

	\item As a second example, consider the flux of $e$-neutrinos induced by the decays $K\to\pi e\nu$ in a collider experiment. Given a known kaon flux, the $e$-neutrino flux is given by
	\begin{eqnarray}
		\Phi_e \, = \, \Phi_K \cdot {\rm BR}(K\to\pi e\nu) \, .
	\end{eqnarray}
	The branching ratio can either be taken from measurement, in which case again no correction is needed, or it can be calculated using two experimental inputs: (i) the kaon lifetime;  (ii) the hadronic matrix element. While the first input is not expected to be affected considerably by nonunitarity, the hadronic matrix element, as measured in semileptonic decays, does inherently include effects of nonunitarity coming from the leptonic matrix elements in the decay (as in Eq.~\eqref{eq:leptonic matrix element}).
	In this case the true flux would differ from the expected SM flux by a ratio of normalization factors.

\end{enumerate}

We conclude that flux estimations either do not need to be corrected for the effects of nonunitarity, or at most are corrected by ratios of normalization factors arising from the product of the leptonic matrix element and measured hadronic matrix element,
\begin{equation}\label{eq: flux normalization}
	\Phi_\alpha = (\Phi_\alpha)^{\rm Predicted}\cdot \frac{N_\alpha}{N_\beta}\, .
\end{equation}
This is in contrast to the widespread treatment of SM flux predictions in the leptonic nonunitarity literature, where a correction factor of $N_\alpha$ is applied to all flux predictions.
We note that in principle, lattice QCD results can be used to estimate hadronic matrix elements. We do not discuss this option further here.
As illustrated in Sec.~\ref{sec:LFU}, ratios of normalization factors, as in Eq.~\eqref{eq: flux normalization}, are constrained to be very close to unity due to LFU constraints, and are therefore negligible when considering ${\cal O}(1)$ unitarity violation.  Additional LFU bounds which are directly related to neutrino production mechanisms are given in Appendix~\ref{app:LFU}.

The situation for interaction cross section predictions depends on the relevant energy scale. 
For low-energy neutrino experiments, below the perturbative regime of QCD, SM predictions generically require experimental hadronic input involving neutrino states, which inherently account for the effects of nonunitarity. In these cases therefore we argue no correction factor should be applied.
For experiments that operate in the perturbative QCD regime, where  the interaction cross section is dominated by deep inelastic scattering (DIS), the situation is subtle and depends on the way input parameters are measured. One can hope that input parameters be measured using processes with no neutrinos.
This is possible for some inputs --- for example, one could imagine that in order to interpret data from collider experiments such as FASER$\nu$~\cite{FASER:2019dxq} and SND@LHC~\cite{SHiP:2020sos}, the weak constant would be taken from LHC measurements rather than from muon decay.
Unfortunately, in the absence of light-quark flavor tagging, the CKM matrix element $|V_{ud}|$ is currently extracted only from low-energy experiments that involve neutrinos, where the most precise method is using super-allowed beta decays~\cite{Hardy:2020qwl}. The combination that is measured introduces one insertion of the normalization, $(G_F^2 |V_{ud}|^2 \Nuniv)$, which cancels out in the DIS cross section prediction.
If one could use only neutrino-free inputs, the definition of SM cross section~\cite{Antusch:2006vwa} would no longer be ambiguous, and a measurement of DIS would be linearly sensitive to $\Nuniv$. 
We conclude that as long as contaminated inputs have to play a part, no nonunitarity correction factor should be applied for the cross section.
Our treatment of flux and cross section predictions, different from what is commonly used in the literature, results in drastically different conclusions as to the impact of neutrino oscillation experiments on nonunitarity; we find that oscillation experiments have no sensitivity to LMM row normalization factors. Below we show what information can be extracted from oscillation experiments under the hypothesis of LMM nonunitarity. %

\subsection{Constraints on $\{|U_{\alpha i}|^2/\Nuniv\}$ and $\Nuniv$  from neutrino oscillation data}
\label{sec:oscillation}
In what follows we perform frequentist statistical analyses, and provide our results in the form of $\Delta\chi^2$ distributions, minimized with respect to all parameters but the parameter of interest. We use best fit values for the mass squared differences $\Delta m_{21}^2 = (7.42 \pm 0.21) \cdot 10^{-5}~{\rm eV}^2 $ and $\Delta m_{31}^2 = (2.52 \pm 0.03) \cdot 10^{-3}~{\rm eV}^2$~\cite{Esteban:2020cvm}.
We emphasize that we do not attempt to perform a global fit to all existing neutrino-related data, but rather use a selection of experimental results that is both straightforward to interpret in terms of a nonunitary LMM, and provides a comprehensive picture of current constraints. Throughout our analysis we neglect matter effects, which we have verified do not significantly change any of our results. From this point on we do not distinguish between noramlization factors of different rows, and only focus on the constraints on the universal normalization, $\Nuniv$.

%
\subsubsection{e-row magnitudes} 
\begin{figure}[t!]
	\centering
	\includegraphics[width=1.\textwidth]{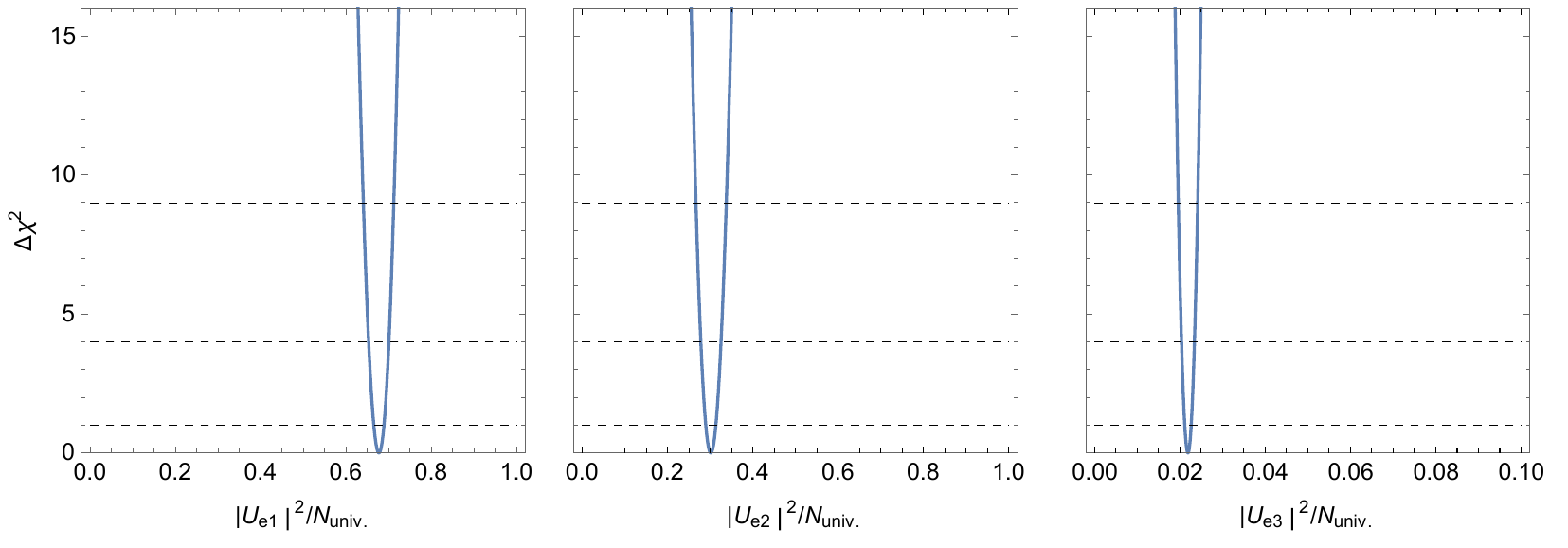}
	\caption{$\Delta\chi^2$ distributions for the normalized $e$-row magnitudes, when taking into account only neutrino oscillation experiments. The expressions and values used to generate these plots are listed in Table~\ref{tab:erow}.}
	\label{fig:erow}
\end{figure}

\bgroup
\def\arraystretch{1.5}
\begin{table}[t]\small
	\begin{center}
		\begin{tabular}{|p{4.6cm}|x{5.8cm}|x{4cm}|}
			\hline
			Experiment & Measured quantity & Value and Ref.  \\
			\hline \hline
			\multirow{2}{*}{Solar CC (SNO + SuperK)} & \multirow{2}{5.5cm}{\centering \large $\frac{|U_{e 2}|^2}{\Nuniv} \left(1 - \frac{|U_{e 3}|^2}{\Nuniv}\right) + \frac{|U_{e3}|^4}{\Nuniv^2}$} & \multirow{2}{*}{$0.293\pm 0.013$~\cite{yasuhiro_nakajima_2020_3959640}} \\
			& & \\
			\hline
			KamLAND &  $4|U_{e1}|^2 |U_{e2}|^2/\Nuniv^2$ &  $0.824\pm 0.036$~\cite{KamLAND:2010fvi}  \\ 
			\hline
			\multirow{2}{*}{Daya Bay} & \multirow{2}{5cm}{\centering \large $\frac{4|U_{e3}|^2}{\Nuniv}\left( 1 -\frac{|U_{e3}|^2}{\Nuniv}\right) $}   & \multirow{2}{*}{$0.0856\pm 0.0029$~\cite{DayaBay:2018yms}}   \\
			& & \\
			\hline \hline
		\end{tabular}
		\caption{Experiments relevant for $e$-row magnitudes and the expressions for the corresponding measured quantities, without assuming unitarity. }
		\label{tab:erow}
	\end{center}
\end{table}
\egroup
Neutrino experiments relevant to the $e$-row (reactor and solar) provide three equations for the three normalized $e$-row parameters, 
\begin{equation}
	\left\{\frac{|U_{e1}|^2}{\Nuniv},\,\frac{|U_{e2}|^2}{\Nuniv},\, \frac{|U_{e3}|^2}{\Nuniv}\right\}\, ,
\end{equation}
and can be analyzed separately from all other constraints.
While in the unitary case the three equations can be solved unambiguously for each of the squared magnitudes, $\{|U_{ei}|^2\}$, when considering nonunitarity only the normalized magnitudes are determined, while the overall normalization remains completely unconstrained,
\begin{equation}
	\Nuniv\, \in [0,1]\, .
\end{equation} 
This result is apparent from the expressions in Table~\ref{tab:erow} which depend only on $|U_{e\alpha}|^2/\Nuniv$.
Fig.~\ref{fig:erow} shows the 1D $\Delta\chi^2$ distributions for the normalized elements. 
%

\subsubsection{$\nu_\mu$-beam oscillation data (NOvA, T2K, NOMAD and OPERA)} 
\bgroup
\def\arraystretch{1.5}
{\setlength{\extrarowheight}{5pt}
\begin{table}[t]\small
	\begin{center}
		\begin{tabular}{|p{3.2cm}|x{6.2cm}|x{3.2cm}|x{4.2cm}|}
			\hline
			Experiment & Measured quantity & Value and Ref. & Notes \\
			\hline \hline
			$\overset{(-)}\nu_\mu\to\overset{(-)}\nu_\mu$ T2K &  \multirow{2}{6cm}{\centering \large $\frac{4|U_{\mu3}|^2}{\Nuniv}\left(1 -\frac{|U_{\mu 3}|^2}{\Nuniv}\right)$}  & $1.01\pm 0.02$~\cite{T2K:2020nqo}  & \multirow{2}{4.2cm}{\footnotesize{Interpreting the measurement of $\sin^2\theta_{23}$ as providing the coefficient of $\sin^2x_{31}$}}  \\
			$\overset{(-)}\nu_\mu\to\overset{(-)}\nu_\mu$ NOvA &  &  $0.99\pm 0.02$~\cite{NOvA:2021nfi} &   \\
			\hline
			\multirow{3}{*}{NOMAD} &  $|t_{\mu e}|^2/\Nuniv^2$& \multicolumn{2}{|c|}{$\leq 1.4\cdot 10^{-3}$ at $90\%$ C.L.~\cite{NOMAD:2003mqg}  }   \\ 
			& $|t_{\mu \tau}|^2/\Nuniv^2$ & \multicolumn{2}{|c|}{$\leq 3.3\cdot 10^{-4} $ at $90\%$ C.L.~\cite{NOMAD:2001xxt} }\\ 
			& $|t_{e \tau}|^2/\Nuniv^2$  & \multicolumn{2}{|c|}{$\leq 1.5\cdot 10^{-2} $ at $90\%$ C.L.~\cite{NOMAD:2001xxt}} \\
			\hline
			\multirow{3}{*}{OPERA} & \multirow{3}{6.2cm}{\centering \footnotesize $P_{\mu\tau}= \big[|t_{\mu\tau}|^2 + 4|U_{\mu 3}|^2 |U_{\tau 3}|^2\sin^2 x_{31} + 4|t_{\mu\tau}| |U_{\mu 3}| |U_{\tau 3}| \sin x_{31} $ $\times  \sin\left(x_{31}+\arg(t_{\mu\tau})-\phi_{\tau 3}\right) \big]/\Nuniv^2$ }&  \multirow{3}{3.2cm}{\centering $(1.14 \pm 0.38)$ $ \times \sin^2 x_{31}^{\rm OPERA}$~\cite{OPERA:2018nar}  } & \multirow{3}{4.2cm}
			{\footnotesize Using  $x_{31}\equiv \Delta m_{31}^2 L/ 4E $,\\ 
				\scalebox{0.85}{$ L = 730\,{\rm km},\, E = 17\,{\rm GeV}$}} \\
			& & & \\ 
			& & &\\
			\hline \hline
		\end{tabular}
		\caption{Select $\nu_\mu$-beam oscillation experiments and the expressions for the corresponding measured quantities, without assuming unitarity. }
		\label{tab:mutau}
	\end{center}
\end{table}
\egroup
Oscillation data from $\overset{(-)}{\nu_\mu}\to\overset{(-)}{\nu_\mu}$ (T2K and NOvA) 
and zero distance nonclosure constraints from the NOMAD experiment, provide information on the normalized parameters,
\begin{equation}
	\left\{\frac{|U_{\mu i}|^2}{\Nuniv},\, \frac{|U_{\tau i}|^2}{\Nuniv}\right\}\, .
\end{equation}
By looking at the expressions of  Table.~\ref{tab:mutau}, one can see that T2K and NOvA~\cite{NOvA:2021nfi}, fix $|U_{\mu 3}|^2/\Nuniv$ to be centered around~$\sim 1/2$, while the NOMAD~\cite{NOMAD:2001xxt} and OPERA measurements provide constraints on the remaining normalized magnitudes. 
The resulting $\Delta\chi^2$ distributions are shown in Fig.~\ref{fig:Umutau}. We note that while appearance experiments can place bounds on the magnitude of nonclosure parameters, $|t_{\alpha\beta}|$, the normalization, $\Nuniv$, remains completely unconstrained by the combination of all existing oscillation data. 

\subsubsection{A final comment on oscillation experiments} 
The results above should come with no surprise. After we argue that fluxes and cross sections should not be rescaled, it is clear that oscillation experiments measure the transition probability. As can be seen from the explicit expressions of Appendix~\ref{app:oscillation probabilities}, the probability expressions are only sensitive to the normalized matrix elements, $|U_{\alpha i}|^2/\Nuniv$. The same is true for the prospects of the DUNE experiment, for which the relevant parameters are once again the normalized magnitudes and the normalized nonclosure factors, $|t_{\alpha\beta}|/\Nuniv$, while no sensitivity exists to the diagonal nonunitarity parameters, embodied by the overall normalization, $\Nuniv$.

One might wonder if matter effects introduce a sensitivity to the overall normalization. We argue that this is not the case. Working in the basis of vacuum mass eigenstates, the matter potential has the form [see e.g.~\cite{Blennow:2016jkn}],
\begin{align}
	H_{\rm mat.} & = U^\dagger \text{diag}\{V_{CC} + V_{NC},\,V_{NC},\,V_{NC}\} U~,
\end{align} 
which introduces a quadratic dependence on the LMM matrix elements. Nevertheless, the matter potential is proportional to the Fermi constant measured from muon decay, as in Eq.~\eqref{eq: Gmu}. As a result, matter effects only introduce additional sensitivity to $|U_{\alpha i}|^2/\Nuniv$, and no net sensitivity to $\Nuniv$ itself.\footnote{If the Fermi constant is measured from a neutrinoless process, such as high energy measurements of the weak coupling at the LHC, then matter effects do introduce sensitivity to $\Nuniv$. We believe this is unintentionally the source of the DUNE sensitivity to the parameter $\alpha_{ee}$ in~\cite{Blennow:2016jkn}. Nevertheless, as the DUNE experiment preforms at energies much closer to those of the muon decay than those of LHC measurements, we do not see any justification for such an assumption.} 

To conclude, within the assumption of kinematically inaccessible extra states, oscillation experiments, including the future DUNE, are only sensitive to the magnitudes of nonclosure factors $|t_{\alpha\beta}|$, which should be the only target of those experiments under the hypothesis of nonunitarity.
Once again, just as in the case of flux and cross section estimation, we see that contamination in input parameters cancels out the sensitivity to the overall normalization.

\begin{figure}[t!]
	\centering
	\includegraphics[width=0.8\textwidth]{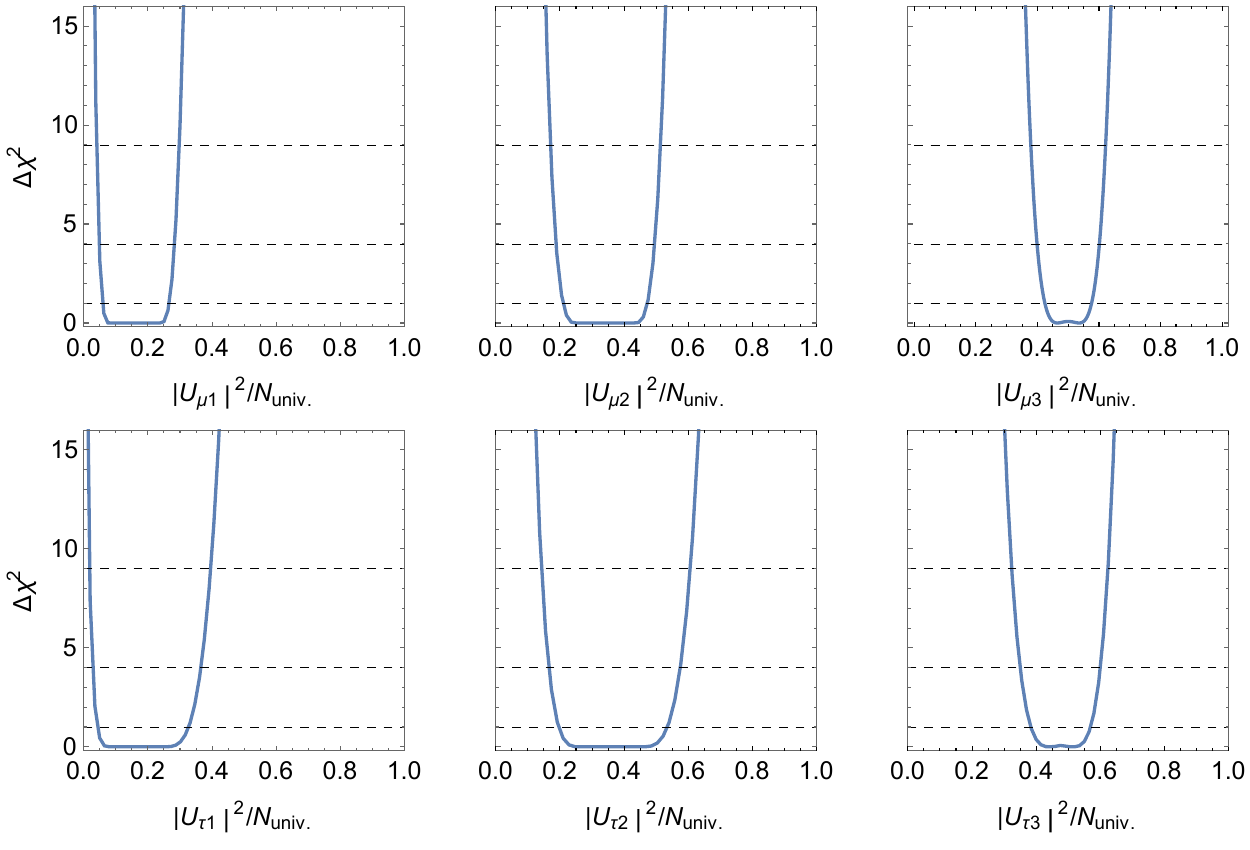}
	\caption{$\Delta\chi^2$ distributions for $\{|U_{\mu i}|^2/\Nuniv,\, |U_{\tau i}|^2/\Nuniv\}$ from muon neutrino beam experiments. }
	\label{fig:Umutau}
\end{figure}
%

\section{Discussion and Conclusions}
\label{sec:conclusion}
%
In this work we have revisited the topic of leptonic nonunitarity resulting from kinematically inaccessible heavy degrees of freedom. While traditionally the literature separates between direct measurements coming from neutrino experiments which are considered as model independent, and a set of all other indirect constraints, we argue that such separation is unjustified, and even inconsistent. 

In our analysis we choose a different differentiation.
In the minimal and robust dataset we include measurements of LFU ratios together with weak-mixing angle constraints. This includes two measurements of LFU ratios, and one overconstraining measurement of the weak mixing angle. This minimal dataset is sufficient to constrain nonunitarity at the $\mathcal{O}(10^{-3})$ level. 
We emphasize that both LFU and the weak mixing angle were measured in many different systems. Moreover, it requires quite remarkable conspiracy between other new physics effects to mimic the SM prediction - hence we term this minimal set as robust. 

Next we revisit the {\it Oscillation $+$ LFU} dataset, and stress that for consistency the LFU bounds must be included when analyzing oscillation data under the hypothesis of neutrino nonunitarity.
In our assessment of the knowledge gained from neutrino oscillation data, we draw special attention to the important role of experimental inputs to fluxes, cross sections and matter effects. While it is traditionally stated that neutrino fluxes and cross sections should be corrected by a normalization factor compared to the SM predictions, we argue that so-called "SM predictions" are contaminated by the same NP leading to nonunitarity, and therefore do not commonly require any correction.

Our different treatment of flux and cross section normalization, as well as the inclusion of LFU bounds, show that oscillation experiments are only sensitive to the normalized matrix elements, $\left|U_{\alpha\beta}\right|^2/N$, and that there is a flat direction in the {\it universal normalization} of the matrix. Once input parameters are treated correctly, this conclusion remains even in the presence of matter effects. 
This result is in sharp contradiction with the traditional treatment of oscillation data in the literature, and holds also for future oscillation experiments. In particular, the DUNE physics book~\cite{DUNE:2020ypp} claims for $\mathcal{O}(10\,\%)$ sensitivity to $\Nuniv \simeq N_e = (1-\alpha_{ee})^2$ , while we find no sensitivity to $\Nuniv$ exists.  

Our study highlights the strength of zero-distance oscillation experiments in providing knowledge of the nonclosure of a nonunitary LMM, by placing bounds on the factors $|t_{\alpha\beta}|/N$. The remarkable bounds placed on $|t_{\mu\tau}|^2/N^2$ by the NOMAD~\cite{NOMAD:2001xxt} and CHORUS~\cite{CHORUS:2007wlo} experiments, of $\mathcal{O}(10^{-4})$, are the only existing bounds from oscillation data that come somewhat close to the level of electroweak and LFU constraints. We believe that improving the NOMAD bounds on the nonclosure factors should be the target of future experiments that wish to constrain LMM nonunitarity. 
We note, that our conclusions hold for the case of kinematically inaccessible heavy neutrino states. Other scenarios that result in a nonunitary LMM  are discussed, for example, in Ref.~\cite{Blennow:2016jkn}.

%
As a last comment, we find it remarkable that the resulting bounds from the minimal and robust dataset are two orders of magnitude stronger then the bounds on CKM nonunitarity.

\subsection*{Acknowledgments}
We thank Yossi Nir for carefully reading the manuscript and for many insightful comments. We thank Yuval Grossman and Yotam Soreq for useful discussions.
The work of D.A. is supported by the U.S. Department of Energy (DOE) under Award No. DE-SC0015845. 
A.D is an awardee of the Women's Postdoctoral Career Development Award.

\bibliographystyle{utphys}
\bibliography{nonU}

\newpage
\appendix

\section{General oscillation probability expressions}
\label{app:oscillation probabilities}
Using Eqs.~(\ref{eq: ket neutrino state},\ref{eq:Prob}), we derive the oscillation probability expressions when allowing for  nonunitarity. Defining 
\begin{equation}
	x_{ij} \, \equiv \, \frac{\Delta m_{ij}^2 L}{4E}\, ,
\end{equation}
we have for the disappearance probability,
\begin{eqnarray}
	P(\nu_\alpha\to\nu_\alpha) \, &=& \,  1 - \frac{4|U_{\alpha 2}|^2(|U_{\alpha 1}|^2 + |U_{\alpha 3}|^2)}{N_\alpha^2}\sin^2 x_{21} - \frac{4|U_{\alpha 3}|^2(|U_{\alpha 1}|^2 + |U_{\alpha 2}|^2)}{N_\alpha^2}\sin^2 x_{31} \\ \nonumber
	&\,& + \frac{8|U_{\alpha 2}|^2|U_{\alpha 3}|^2}{N_\alpha^2}\sin x_{21}\sin x_{31}\cos x_{32}\, .
\end{eqnarray}
The appearance probability expression is similarly given by
\begin{eqnarray}
	P(\overset{(-)}{\nu_\alpha}\to \overset{(-)}{\nu_\beta}) \, & = & \, \frac{|t_{\alpha\beta}|^2}{N_\alpha N_\beta } + \frac{4|U_{\alpha 2}|^2|U_{\beta 2}|^2}{N_\alpha N_\beta}\sin^2x_{21} + \frac{4|U_{\alpha 3}|^2|U_{\beta 3}|^2}{N_\alpha N_\beta}\sin^2x_{31} \\ \nonumber
	&\,& + \frac{8|U_{\alpha 2}||U_{\beta 2}||U_{\alpha 3}||U_{\beta 3}|}{N_\alpha N_\beta} \sin x_{21} \sin x_{31} \cos\left(x_{32} \mp \Delta\phi_2^{\alpha\beta} \pm \Delta \phi_3^{\alpha\beta}	\right) \\ \nonumber
	&\,& + \frac{4|t_{\alpha\beta}|}{N_\alpha N_\beta}\Bigg[ |U_{\alpha 2}| |U_{\beta 2}|\sin x_{21} \sin\left(x_{21} \pm \arg(t_{\alpha\beta})\pm \Delta\phi_2^{\alpha\beta}\right) \\ \nonumber
	 &\,&\quad  +   |U_{\alpha 3}| |U_{\beta 3}|\sin x_{31} \sin\left(x_{31} \pm \arg(t_{\alpha\beta}) \pm \Delta\phi_3^{\alpha\beta}\right)\Bigg]\, ,
\end{eqnarray}
where $ \Delta\phi_{i}^{\alpha\beta} \, \equiv \, \phi_{\alpha i}-\phi_{\beta i}$.
	
To derive the expression for solar neutrinos, we use $\varphi_{ei}$ - the outgoing fractional fluxes of neutrino mass eigenstates  leaving the sun. Importantly, as a consequence of the neutrino mass eigenstates being orthonormal, $\sum \varphi_{ei}=1$ . Taking matter effects into account, the fractional fluxes are given by $ \varphi_{ei} = \{0, \, 1- \left|U_{e3}\right|^2/\Nuniv, \, \left|U_{e3}\right|^2/\Nuniv \}$~(see e.g., Ref.~\cite{Ellis:2020hus}), and the probability of measuring $e$-neutrino on earth is given by  $P_{ee} =    \varphi_{ei}  P_{ie} = \varphi_{ei}  \left|{U}_{ei}\right|^2 / \Nuniv $. %

\section{LFU bounds from neutrino production processes}
\label{app:LFU}
\begin{table}[t]
	\begin{center}
		\bgroup
		\def\arraystretch{1.}
		\begin{tabular}{|c||c |c| c |c |}
			\hline
			Ratio &  $\ell_\alpha \to \ell_\beta \nu_\alpha\bar{\nu}_\beta$ & $\pi \to \ell_\alpha \bar{\nu}_\alpha$ & $K \to \ell_\alpha \bar{\nu}_\alpha$ & $D_s \to \ell_\alpha \bar{\nu}_\alpha$\\ 
			\hline \hline
			$\frac{N_\mu}{\Nuniv}$ & $1.004 \pm 0.003$ ~\cite{HFLAV:2019otj} & $0.999 \pm 0.002$~\cite{PiENu:2015seu} & $1.004 \pm 0.013$~\cite{NA62:2011aa}& $-$\\
			\hline
			$\frac{N_\tau}{\Nuniv}$ & $1.006 \pm 0.003$~\cite{HFLAV:2019otj} & $-$ & $-$ &$-$\\
			\hline
			$\frac{N_\tau}{N_\mu}$ &  $1.002 \pm 0.003$~\cite{HFLAV:2019otj}& $0.992\pm 0.005 $~\cite{HFLAV:2019otj} & $0.996\pm 0.013 $~\cite{HFLAV:2019otj} & $1.016 \pm 0.035$~\cite{ParticleDataGroup:2020ssz}\\
			\hline
		\end{tabular}
		\egroup
		\caption{Leading bounds on LFU which are directly related to the neutrino source of various experiments. For accelerator and collider neutrino experiments, electron neutrino are dominantly produced in Kaon decay, $\mu$-neutrinos are mainly produced from both pion and kaon decay, and $\tau$-neutrinos are mainly produced from $D_s$ decays.}
		\label{tab:LFU2}
	\end{center}
\end{table}

The LFU constraints are inherently related to the bounds from neutrino oscillation experiments, and some processes are actually exactly the source processes for atmospheric, accelerator and collider neutrino experiments. Let us illustrate it explicitly  by using pion decay as an example. Consider a neutrino state Eq.~\eqref{eq: ket neutrino state}
\begin{align}
	\ket{\nu_\alpha^{\rm eff.} (t)} & = \frac{1}{\sqrt{N_{\alpha}}}\sum_{k=1}^{3} U^*_{\alpha k} e^{i E_k t} \ket{\nu_k}~,
\end{align}
as produced from pion decay. At the source, $t=0$, the ratio of branching ratio of pion decaying to a muon and neutrino, compared to that of pion decaying to electron and neutrino is given by (the $\tau$ bounds are similar and coming from $\tau \to \pi \nu$)
\begin{align}\label{eq:LFU_example_normalization}
	\frac{{\rm BR}(\pi \to \mu \nu)}{{\rm BR}(\pi \to e \nu)} & = R_{\mu e} \frac{\left|\braket{ \mu^+ \nu_\mu^{\rm eff.}| \mathcal{H} |\pi^+ }\right|^2}{\left|\braket{e^+ \nu_e^{\rm eff.} | \mathcal{H} | \pi^+}\right|^2}~,
\end{align}
where $R_{\mu e}$ is a kinematical factor which to leading order is equal to $m_\mu^2 (m_\pi^2 - m_\mu^2)^2 / m_e^2 (m_\pi^2 - m_e^2)^2$. Plugging the neutrino states defined above, and taking into account the rotation of the interaction in the Hamiltonian as we move from the interaction basis to the mass basis, one finds
\begin{align}
	\frac{{\rm BR}(\pi \to \mu \nu)}{{\rm BR}(\pi \to e \nu)} R_{\mu e}^{-1}& =  \frac{\left|\frac{1}{\sqrt{N_\mu}}\sum_{i=1}^{3} \left|U_{\mu i}\right|^2\right|^2}{\left|\frac{1}{\sqrt{\Nuniv}}\sum_{i=1}^{3} \left|U_{e i}\right|^2\right|^2} = \frac{N_\mu}{\Nuniv}~.
\end{align}
In other words, under the assumption of no conspiracy of BSM invisible light states, using the neutrino states Eq.~\eqref{eq:  ket neutrino state} as produced by a pion source, is directly related to measuring LFU from pion decay, which bounds the ratio of the normalization factor. 
In Table~\ref{tab:LFU2} we summarize LFU constraints which are directly related to the production of neutrino beams. We note that nonuniversality in $W$-boson decays also places constraints, but these are slightly weaker and are relevant at a different physical scale, therefore we do not include them here.

\end{document}